\documentclass[aps,reprint,showpacs,nofootinbib,superscriptaddress,floatfix]{revtex4-1}
\usepackage{amsmath,latexsym,amssymb,hyperref,graphicx}
\usepackage{times}
\usepackage{epsfig}
\usepackage{slashed}

\hypersetup{colorlinks,citecolor= nicegreen,linkcolor= nicered}
\usepackage{color}
\definecolor{nicered}{rgb}{0.7,0.1,0.1}
\definecolor{nicegreen}{rgb}{0.1,0.5,0.1}

\newcommand{\be}{\begin{equation}}
\newcommand{\ee}{\end{equation}}
\newcommand{\bea}{\begin{eqnarray}}
\newcommand{\eea}{\end{eqnarray}}
\def\<{\langle}
\def\>{\rangle}

\newcommand\SEC[1]{\medskip\noindent{\sl\bfseries #1}}

\begin{document}

\title[ Two-loop Wess-Zumino model with exact supersymmetry on the lattice]{
Two-loop Wess-Zumino model with exact supersymmetry on the lattice}

\author{Alessandra \surname{Feo}}
\affiliation{Dipartimento di Fisica e Scienze della Terra, 
Universit\`a di Parma, \\ 
Viale G.P. Usberti 7/A. I-43124 Parma, Italy} 

\date{\today} 

\begin{abstract} 
We consider a lattice formulation of the four dimensional $N=1$ 
Wess-Zumino model in terms of the Ginsparg-Wilson relation.
This formulation has an exact supersymmetry on the lattice.
The lattice action is invariant under a deformed supersymmetric
transformation which is non-linear in the scalar fields and it 
is determined by an iterative procedure 
in the coupling constant to all orders in perturbation theory.
We also show that the corresponding Ward-Takahashi identity
is satisfied at fixed lattice spacing. The calculation is 
performed in lattice perturbation theory up to order $g^3$
(two-loop) and the Ward-Takahashi identity (containing 
110 connected non-tadpole Feynman diagrams)
is satisfied at fixed lattice spacing thanks to this exact 
lattice supersymmetry.

\end{abstract}


\maketitle

\SEC{Introduction.}
Non-perturbative dynamics play an important role in the theory
of supersymmetry breaking needed in order to produce a low-energy
four-dimensional effective action with a residual $N=1$ supersymmetry.
For this reason, much effort has been dedicated to
formulating a lattice version of supersymmetric theories
(see
\cite{Catterall:2010jh,Catterall:2009it,Giedt:2009yd,Feo:2004kx,Kaplan:2003uh,Feo:2002yi}).

The major obstacle in formulating a supersymmetric theory on the lattice is 
that the supersymmetry is a part of the super Poincar\'e group, which is 
explicitly broken by the lattice. Ordinary Poincar\'e invariance is 
also broken by the lattice, but the operators that violate Poincar\'e
symmetry are all irrelevant (i.e., goes to zero in the continuum limit, 
$a \to 0$). In the case of a supersymmetric theory, these operators are 
relevant and a fine tuning is needed in order to eliminate their contribution.
This is the case of the Wilson fermion approach for the $N=1$ super Yang-Mills 
theory, in which the only operator that violates the $N=1$ supersymmetry is a 
fermion mass term \cite{Montvay:2001aj}. By tuning the fermion mass to the 
supersymmetric limit, one recovers supersymmetry in the continuum limit 
(see for example Ref.~\cite{Farchioni:2001wx,Feo:1999hw,Demmouche:2010sf,Bergner:2013nwa}). 
Alternatively, using domain wall fermions \cite{Kaplan:1999jn,Fleming:2000fa}
this fine tuning is not required. See also Ref.~\cite{Giedt:2008xm,Endres:2009yp}.
This is in contrast with lower dimensional models (with extended supersymmetry)
where the lattice symmetries can eliminate the need for such a fine tuning.
Basically, the strategy here is to realize part of the supercharges as an exact symmetry 
on the lattice. This exact supersymmetry is expected to play a key role to restore  
continuum supersymmetry without (or with less) fine tuning 
\cite{Catterall:2001fr,Catterall:2004np,Kaplan:2002wv,D'Adda:2005zk,Arianos:2008ai,D'Adda:2010pg,Beccaria:2004ds}. 

We consider the $N=1$ four dimensional lattice Wess-Zumino model 
introduced in Refs.~\cite{Fujikawa:2001ns,Fujikawa:2002ic,Fujikawa:2001ka} 
and studied in \cite{Bonini:2004pm,Bonini:2005qx} (for a numerical approach
see Refs.~\cite{Chen:2010uca,Giedt:2010zz,Chen:2011uea}). Although it is a toy model,
all the difficulties of lattice supersymmetry are already present.
A necessary condition to have exact lattice supersymmetry is that the 
associated Ward-Takahashi identity (WTi) has to be exactly satisfied 
on the lattice. That exact symmetry is responsible for the restoration
of supersymmetry in the continuum limit without fine tuning of the parameters
of the action.

Here we extend the formulation introduced in \cite{Bonini:2004pm} and show
that it is possible to define a deformed lattice supersymmetric
transformation which leaves the full action invariant at fixed lattice
spacing, to all orders in perturbation theory.
This transformation is nonlinear in the scalar fields. The action 
and the transformation are written in terms of the Ginsparg-Wilson
operator and reduce to their continuum expression in the naive continuum limit
\cite{Bonini:2004pm}.
Since in presence of any exact symmetry all the WTi are fulfilled,
we did checked that the simplest non trivial one, i.e.,
the one-point WTi, is exactly satisfied on the lattice for both, 
one-(order $g$) and two-(order $g^3$) loop.
Although, in a one-point WTi calculation the order $g^3$ is a non trivial zero,
it shows cancellations between fermion and scalar fields contributions as
required by the supersymmetry. 
This result extend to two-loop order the results already obtained in
\cite{Bonini:2005qx} for a different WTi, i.e., the one loop (two-points) WTi (order $g^2$). 
In this case, the exact lattice supersymmetry  
determines the finite part of the scalar and fermion renormalization wave function
which coincide in the continuum limit and leads to the restoration of the  
continuum supersymmetry.

In the following, we briefly review 
the $N=1$ four dimensional lattice Wess-Zumino model based on the
Ginsparg-Wilson fermion operator, and shows how to build up a lattice
supersymmetry transformation which is an exact symmetry of the lattice
action, to all orders in perturbation theory. 
In the remaining part, we derive the WTi and explicitly check 
that the one-point WTi up to two loop is exactly satisfied at fixed lattice spacing. 

\SEC{The lattice Wess-Zumino model.} 
\label{wz}
We formulate the lattice Wess-Zumino model in four dimensions
introducing a Dirac operator $D$ that satisfies
the Ginsparg-Wilson relation \cite{Ginsparg:1981bj},
\be
\gamma_5 D + D \gamma_5 = a D \gamma_5 D \, .
\label{gw}
\ee
This relation implies a continuum symmetry of the fermion action
which may be regarded as a lattice form of the chiral symmetry
\cite{Luscher:1998pqa,Hernandez:1998et} and protects the 
fermion masses from additive renormalization.
Although our analysis is valid for all $D$'s that satisfy 
Eq.~(\ref{gw}), we use the solution given by
Neuberger \cite{Neuberger:1997fp,Neuberger:1998wv},
\begin{eqnarray}
D &=& \frac{1}{a} (1 - X \frac{1}{\sqrt{X^\dagger X} }) \, ,
 \qquad X = 1 - a D_w \, ,
\label{D}
\\
D_w &=& \frac{1}{2} \gamma_\mu ( \nabla^\star_\mu + \nabla_\mu ) -
\frac{a}{2} \nabla^\star_\mu \nabla_\mu
\label{Dw}
\end{eqnarray}
and $\nabla_\mu \phi(x)$ and $\nabla_\mu^\star \phi(x)$, are 
the forward and backward lattice derivatives, respectively.
Plugging Eq.~(\ref{Dw}) in (\ref{D}) we find it convenient to isolate in
$D$ the part containing the gamma matrices \cite{Bonini:2004pm}, and 
write $D$ as,
$D = D_1 + D_2$, where
$D_1 = \frac{1}{4} \mbox{Tr}(D)$
and
$D_2 = \frac{1}{4} \gamma_\mu \mbox{Tr}(\gamma_\mu D)$.
More explicitly we have,
\begin{eqnarray}
D_1 = \frac{1}{a} [ 1 - (1 + \frac{a^2}{2} \nabla^\star_\mu \nabla_\mu)
\frac{1}{\sqrt{X^\dagger X}} ] \, ,
\label{d1} \\
D_2 = \frac{1}{2} \gamma_\mu (\nabla^{\star}_\mu + \nabla_\mu)
\frac{1}{\sqrt{X^\dagger X}} \equiv \gamma_\mu D_{2 \mu} \, .
\label{d2}
\end{eqnarray}
In terms of $D_1$ and $D_2$ the Ginsparg-Wilson relation (\ref{gw}) 
becomes \cite{Bonini:2004pm},
\be
D_1^2 - D_2^2 = \frac{2}{a} D_1 \, , ~~~
\mbox{and} ~~~
(1-\frac a2 D_1)^{-1}D_2^2=-\frac 2a D_1 \, .
\label{gw1}
\ee
The action of the 4-dimensional Wess-Zumino model on the lattice was
introduced in Refs.~\cite{Fujikawa:2001ns,Fujikawa:2002ic,Fujikawa:2001ka}. 
In our notation,
\begin{align}
S_{WZ} &= \sum_x \{ \frac{1}{2} \bar \chi (1 - \frac{a}{2} D_1)^{-1}
D_2 \chi - \frac{2}{a} \phi^{\dagger}D_1 \phi  \nonumber \\
& + F^{\dagger} (1 - \frac{a}{2} D_1)^{-1} F + \frac{1}{2} m \bar \chi \chi
+ m (F \phi + (F \phi)^\dagger) \nonumber \\
& + g \bar \chi (P_{+} \phi P_{+} +
P_{-} \phi^{\dagger} P_{-}) \chi + g (F \phi^2 + (F \phi^2)^\dagger) \} \, ,
\nonumber
\end{align}
where $\phi$ and $F$ are scalar fields and $\chi$ is a Majorana fermion that
satisfies the Majorana condition: $\bar \chi = \chi^T C $ 
and $C$ is the charge conjugation matrix that satisfies:
$C^T = -C$ and $C C^\dagger = 1$. 
Moreover, 
$C \gamma_\mu C^{-1} = - (\gamma_\mu)^T$ and 
$C \gamma_5 C^{-1} = (\gamma_5)^T$.
In the continuum limit, i.e., $a \to 0$, $S_{WZ}$ reduces to the 
continuum Wess-Zumino action,
\begin{align}
S & = \int \{ \frac{1}{2} \bar \chi (\not \partial + m) \chi +
\phi^\dagger \partial^2 \phi +  F^{\dagger} F +
m (F \phi + (F \phi)^\dagger) \nonumber \\
& + g \bar \chi (P_{+} \phi P_{+} + P_{-} \phi^{\dagger} P_{-}) \chi +
g (F \phi^2 + (F \phi^2)^\dagger) \} \, . \nonumber
\end{align}

If one defines the real components by
$\phi \to \frac{1}{\sqrt{2}} (A + i B)$ and 
$F \to \frac{1}{\sqrt{2}} (F - i G)$, 
the Wess-Zumino action can be written as a free part or kinetic term, 
$S_0$, plus the interaction term, $S_{int}$, 
$S_{WZ} = S_0 + S_{int}$, as,
\begin{align}
S_{0} & = \sum_x \{ \frac{1}{2} \bar \chi
(1 - \frac{a}{2} D_1)^{-1} D_2 \chi - \frac{1}{a} (A D_1 A + B D_1 B) 
\nonumber \\
& + \frac{1}{2} F (1 - \frac{a}{2} D_1)^{-1}F + 
\frac{1}{2} G (1 - \frac{a}{2} D_1)^{-1} G \}  \, , \nonumber \\
S_{int} & = \sum_x \{ \frac{1}{2} m \bar \chi \chi + m (F A + G B) 
\nonumber \\
& + \frac{1}{\sqrt{2}} g \bar \chi (A + i \gamma_5 B) \chi 
+ \frac{1}{\sqrt{2}} g [ F (A^2 - B^2 ) + 2 G (A B) ] \} \nonumber
\end{align}
and from here we get the propagators for the scalar 
and fermion fields:
\begin{align}
& \< A A \> = \< B B \> = -{\cal M}^{-1} {\mathbb D}_1^{-1}  \nonumber \\
& \< F F \> = \< G G \> = \frac{2}{a}{\cal M}^{-1} D_1
= -{\cal M}^{-1} {\mathbb D}_1^{-1} D_2^2  \nonumber \\
& \< A F \> = \< B G \> = m \, {\cal M}^{-1}  \nonumber \\
& \< \chi \bar \chi \> =
({\mathbb D}_1^{-1} D_2 + m)^{-1} = 
 -{\cal M}^{-1} ({\mathbb D}_1^{-1} D_2 - m) \, ,
\label{prop}
\end{align}
where 
$ {\mathbb D}_1^{-1} \equiv (1 - \frac{a}{2} D_1)^{-1}$
and 
${\cal M}^{-1} \equiv [ \frac{2}{a} D_1 (1 - \frac{a}{2} D_1)^{-1} 
+ m^2 ]^{-1}$
and the Ginsparg-Wilson relation (\ref{gw1}) has been used to rewrite the
auxiliary fields propagators.
Despite the appearance of the operator $(1-\frac{a}{2}D_1)^{-1}$, there are
no would be doublers and the propagators are regular (see appendix~A of 
Ref.~\cite{Bonini:2005qx} for details). For a non-perturbative approach
that shows localization of this operator,
see Refs.~\cite{Chen:2010uca,Giedt:2010zz}.

\SEC{The supersymmetric transformation.}
\label{trans}
As was discussed in \cite{Fujikawa:2001ns} and then shown in 
\cite{Bonini:2004pm}, $S_0$ is invariant under a lattice supesymmetric
transformation,
\begin{align}
 \delta A & = \bar \varepsilon \chi = \bar \chi \varepsilon \, , 
~~~ \delta B = -i \bar \varepsilon \gamma_5 \chi = -i \bar \chi \gamma_5
\varepsilon \nonumber \\
 \delta \chi & = - D_2 (A - i \gamma_5 B) \varepsilon - (F - i \gamma_5 G)
\varepsilon \nonumber \\
 \delta F & = \bar \varepsilon D_2 \chi \, , 
~~~ \delta G = i \bar \varepsilon  D_2 \gamma_5 \chi \, ,
\label{susytransf}
\end{align}
which is similar to the continuum one except for replacing the continuum derivative 
with the lattice one, $D_{2\mu}$.
Indeed, \cite{Bonini:2004pm} the variation of $S_0$ under this transformation
is
\begin{align}
\delta S_0 & = \sum_x \{ \bar \chi (1 - \frac{a}{2} D_1)^{-1} D_2 
[-D_2 (A - i \gamma_5 B) \varepsilon - \nonumber \\ 
& (F - i \gamma_5 G) \varepsilon ] - \frac{2}{a} \bar \chi \varepsilon D_1 A
+ \frac{2 i}{a} \bar \chi \gamma_5 \varepsilon D_1 B +  \nonumber \\
& (\bar \varepsilon D_2\chi) (1 - \frac{a}{2} D_1)^{-1} F 
+ i (\bar \varepsilon D_2 \gamma_5 \chi) (1 - \frac{a}{2} D_1)^{-1} G \}
\, . \nonumber
\end{align}
Using the Ginsparg-Wilson relation (\ref{gw1})
and integrating by part (details in \cite{Bonini:2004pm}) we get:
\begin{align}
& \sum_x  \{ -\bar \chi \varepsilon [ (1 - \frac{a}{2} D_1)^{-1} D_2^2
+ \frac{2}{a} D_1 ] A + i \bar \chi \gamma_5 \varepsilon [ 
\nonumber \\
& (1 - \frac{a}{2} D_1)^{-1} D_2^2 + \frac{2}{a} D_1 ] B 
- \bar \chi (1 - \frac{a}{2} D_1)^{-1} D_2 (F - i \gamma_5 G) \varepsilon 
\nonumber \\
& + \bar \chi D_2 \varepsilon (1 - \frac{a}{2} D_1)^{-1}  F
+ i \bar \chi D_2 \gamma_5 \varepsilon (1 - \frac{a}{2} D_1)^{-1} G \}
= 0 \, . \nonumber
\end{align}

As discussed in \cite{Bonini:2004pm,Bonini:2005qx},
the variation of $S_{int}$ under the supersymmetric transformation 
(\ref{susytransf}) does not vanish due to the
failure of the Leibniz rule at finite lattice spacing \cite{Dondi:1976tx}.
In order to discuss the symmetry properties of the lattice Wess-Zumino model
one possibility would be to modify the action by adding irrelevant terms
that make the full action invariant \cite{Bergner:2012nu}. 
Another possibility is to modify the supersymmetric transformation (\ref{susytransf})
so that $S_{int}$ has an exact symmetry for $a \neq 0$. We shall see 
that this procedure is only possible if we use fermions
that satisfy Eq.~(\ref{gw}).
Since the transformation (\ref{susytransf}) leaves invariant $S_0$, the modification
should vanish for $g=0$. The supersymmetric transformation that leaves invariant
$S_{WZ}$ is similar to Eq.~(\ref{susytransf}) where the only difference
is in the variation of the fermion field \cite{Bonini:2004pm},
\be
\delta \chi = -D_2 (A - i \gamma_5 B) \varepsilon - (F - i \gamma_5 G)                                            
\varepsilon + g R \varepsilon
\label{complete} 
\ee
where $R$ is a function to be determined order by order in perturbation
theory imposing the condition $\delta S_{WZ} = 0$.
Expanding $R$ in power of $g$ \cite{Bonini:2004pm}:
\be
R = R^{(1)} + g R^{(2)} + g^2 R^{(3)} + \cdots
\label{expansion}
\ee
and imposing the symmetry condition order by order in perturbation theory
we find \cite{Bonini:2004pm}
\be
R^{(1)} = ((1 - \frac{a}{2} D_1)^{-1} D_2 + m )^{-1} \Delta L
\label{r1}
\ee
where
\begin{align}
\Delta L & \equiv 1/\sqrt{2} \{2 (A D_2 A - B D_2 B) -
D_2 (A^2 - B^2) \nonumber \\ 
& + 2 i \gamma_5 [(A D_2 B + B D_2 A) - D_2 (A B) ] \} \, .
\label{deltaL}
\end{align}
For $n \geq 2$
\be
R^{(n)} = -\sqrt{2} ((1 - \frac{a}{2} D_1)^{-1} D_2  + m)^{-1}
(A + i \gamma_5 B) R^{(n-1)}  \, . \nonumber 
\label{rn}
\ee
Notice that the operator $((1 - \frac{a}{2} D_1)^{-1} D_2  + m)^{-1}$ is 
the free fermion propagator and Eq.~(\ref{complete}), as the function $R$, 
are non-linear in the scalar fields.
Inserting these results in Eq. (\ref{expansion}), the function $R$ to be used
in Eq. (\ref{complete}) 
resumed to all order in pertubation theory is :
\be
R = [(1 - \frac{a}{2} D_1)^{-1} D_2  + m  + \sqrt{2} g (A + i \gamma_5 B) ]^{-1} \,
    \Delta L \, .
\label{dl}
\ee
Thanks to the Ginsparg-Wilson relation (encoded in (\ref{gw1}))   
we were able to resummed up Eq.~(\ref{expansion}) to obtain Eq.~(\ref{dl}),
which contains all orders in perturbation theory and is a closed form that can 
be used for numerical simulations.
In the limit $a \to 0$ Eq.~(\ref{complete}) becomes (\ref{susytransf})
since $\Delta L$ vanishes in this limit. $\Delta L$ is different from zero
due to the breaking of the Leibniz rule at finite lattice spacing.
This resummation would not have been possible using Wilson's fermions 
\cite{Bartels:1982ue}.
Now we want to show that the full Wess-Zumino action, $S_{WZ}$, is invariant
under the supersymmetric tranformation (with Eq.~(\ref{complete})) and 
include all orders in perturbation theory. Indeed, its variation is 
\begin{align}
\delta S_{WZ} & = \sum_x \{ g \bar \chi [(1 - \frac{a}{2} D_1)^{-1} D_2
R + m R ] \varepsilon  \nonumber \\
& - \frac{g}{\sqrt{2}} [ 2 \bar \chi (A + i \gamma_5 B) D_2 (A - i \gamma_5 B)
 \varepsilon \nonumber \\
& - \bar \chi D_2 (A - i \gamma_5 B)^2 \varepsilon ]
+ \sqrt{2} g^2 \bar \chi (A + i \gamma_5 B) R \varepsilon \} \, . \nonumber
\end{align}
Using Eq.(\ref{dl}) after some algebra we get, indeed zero:
\begin{align}
\delta S_{WZ} & = \sum_x \{ g \bar \chi \Delta L \varepsilon - \frac{g}{\sqrt{2}}
[ 2 \bar \chi (A + i \gamma_5 B) D_2 (A - i \gamma_5 B) \nonumber \\
& - \bar \chi D_2 (A - i \gamma_5 B)^2 \varepsilon ] \} = 0 \, . \nonumber
\end{align}

\SEC{One-point Ward-Takahashi identity to two-loop.}
\label{wtig1}
Before going to two loop (order $g^3$) we first want to show how to 
obtain a WTi. This will be useful for the more involved calculation
at two loop.
The WTi is derived from the generating functional which is given by,
$Z[ \Phi, J ] = \int {\cal D} \Phi \exp{-(S_{WZ} + S_J)}$,
where $S_J$ is the source term,  
$S_J = \sum_x J_\Phi \cdot \Phi
\equiv \sum_x \{ J_A \, A + J_B \, B + J_F \, F + J_G \, G + 
\bar \eta \chi \}$.
Using the invariance of both, the Wess-Zumino action and the measure with 
respect to the lattice supersymmetric transformation (\ref{complete}), 
the WTi reads,
$\< J_\Phi \cdot \delta \Phi \>_J = 0$,
where $\delta\Phi$ is given in (\ref{complete}).

We start with the simplest (one-point) WTi which is generated by taking 
the derivative with respect to $\bar \eta$ and setting to zero all 
the sources, that is:
\be
\< D_2 (A - i \gamma_5 B) \> + \< F \> - i \gamma_5 \< G \> - g \< R \> = 0 \, .
\label{WT}
\ee
The order $O(g)$ of this WTi is given by:
\be
\<D_2 (A - i \gamma_5 B) \>^{(1)} + \< F \>^{(1)} - i \gamma_5 \< G \>^{(1)} 
- g \< R^{(1)} \>^{(0)} = 0 \, ,
\label{wt1}
\ee
where the notation $\< {\cal O} \>^{(n)}$ indicates the $n-$order (in $g$) 
contribution to the expectation value of ${\cal O}$.
Using Eq.~(\ref{prop}) it is easy to see that this WTi is satisfied, which
means that when we insert all the terms into the WTi (\ref{wt1}), 
the result is zero (notice that $\<AA\>=\<B B\>$ and $\<AF\>=\<BG\>$).
For instance,
\begin{align}
 \<D_2 & A_x \>^{(1)} = \frac{g}{\sqrt{2}} D_{2xy} [
\<A_y F_u\> (\<A_u A_u\> - \<B_u B_u\> ) \nonumber \\
& + 2 \<A_y A_u\> ( \<A F\>_u + \<B G\>_u - \frac{1}{2} \mbox{Tr} 
\<\bar \chi \chi \>_u ) ] = 0 \, . \nonumber 
\end{align}
Similarly, $\< G \>^{(1)} = 0$ because of $Tr(\gamma_{\mu})=0$ and
\begin{align}
 \< F_x & \>^{(1)} = \frac{g}{\sqrt{2}} [
\< F_x F_u \> ( \< A A \>_u - \< B B \>_u )  \nonumber \\
& + 2 \< F_x A_u \> ( \< A F \>_u + \< B G \>_u - \frac{1}{2} \mbox{Tr} 
\< \chi_u \bar \chi_u \> ) ] = 0 \, . \nonumber 
\end{align}
Finally, the term including $R$ is given by,
\begin{align}
g \< R_x^{(1)} & \>^{(0)} = 
g \< \chi_u \bar \chi_u \> (2 D_{2 yl} \< A_y A_l \> 
- 2 D_{2 yl} \< B_y B_l \> \nonumber \\
& - D_{2 yl} \< A A \>_l + D_{2 yl} \< B B \>_l) = 0 \, . \nonumber 
\end{align}

Now we are ready to verify that the two loop order, $g^3$, in a one-point 
WTi is satisfied at fixed lattice spacing. The calculation is not trivial
and contains 110 connected non tadpole Feynman diagrams 
(from $F$ and the operator $R$). The tadpole contributions cancel out
separately. We start from Eq.~(\ref{WT}): 
\begin{align}
 \<D_2(A - & i \gamma_5 B)\>^{(3)} + \< F \>^{(3)} - i \gamma_5 \< G \>^{(3)} 
- g \< R^{(1)} \>^{(2)} \nonumber \\
& - g^2 \< R^{(2)} \>^{(1)} - g^3 \< R^{(3)} \>^{(0)}  = 0 \, .
\label{wtg3}
\end{align}
The first term of this WTi is zero because of the $\delta$-momentum
conservation and $D_2(k=0)=0$. Also $i \gamma_5 \< G \>^{(3)}$
is trivially zero. Then, one is left with,
\be
\< F \>^{(3)} - g \< R^{(1)} \>^{(2)}
- g^2 \< R^{(2)} \>^{(1)} - g^3 \< R^{(3)} \>^{(0)}  = 0 \, .
\label{left}
\ee
To calculate the expectation value of $F$ one has to insert the interaction
term until order $g^3$, as,
\begin{align}
\< F_x & \>^{(3)} = g^3/(2 \sqrt{2}) \< F_x \{ \nonumber \\
& [ (\chi (A + i \gamma_5 B) \bar \chi)_u + (F (A^2 - B^2) + 2 GAB)_u ]
\nonumber \\
& [ (\chi (A + i \gamma_5 B) \bar \chi)_v + (F (A^2 - B^2) + 2 GAB)_v ]
\nonumber \\
& [ (\chi (A + i \gamma_5 B) \bar \chi)_w + (F (A^2 - B^2) + 2 GAB)_w ]
\}  \>^{(0)} \nonumber 
\end{align}
where $u,v,w$ are dummy indices.
For the remaining terms of the WTi (\ref{left}) involving the operator $R$,
we have:
\begin{align}
g & \< R_x^{(1)} \>^{(2)} = g^3/2 \< \chi \bar \chi\>_{xy} 
\< \Delta L_y \{ \nonumber \\
& [ (\chi (A + i \gamma_5 B) \bar \chi)_u + (F (A^2 - B^2))_u + 2 (GAB)_u ]
\nonumber \\ 
& [ (\chi (A + i \gamma_5 B) \bar \chi)_v  + (F (A^2 - B^2))_v + 2 (GAB)_v ]
 \}  \>^{(0)} \, , \nonumber 
\end{align}
where $\Delta L$ is given in Eq.~(\ref{deltaL}) and $R^{(1)}$ in (\ref{r1}). 
The second term $R^{(2)}$ is given by,
\begin{align}
g^2 & \< R_x^{(2)} \>^{(1)} = - g^3 \< ({\mathbb D}_1^{-1}D_2 + m)^{-1}_{xz} 
(A_z + i \gamma_5 B_z) \nonumber \\  
& ({\mathbb D}_1^{-1} D_2 + m)^{-1}_{zu} \Delta L_w
 [ (\chi (A + i \gamma_5 B) \bar \chi)_w \nonumber \\ 
& + (F (A^2 - B^2))_w + 2 (GAB)_w ] \>^{(0)} \, .  \nonumber
\end{align}
The last term, $R^{(3)}$, is given by,
\vspace{-1mm}
\begin{align}
 g^3 & \< R_x^{(3)} \>^{(0)} = 2 g^3 \< ({\mathbb D}_1^{-1} D_2 + m)^{-1}_{xy} 
(A_y + i \gamma_5 B_y) \nonumber \\ 
& ({\mathbb D}_1^{-1} D_2 + m)^{-1}_{yz} (A_z + i \gamma_5 B_z) 
({\mathbb D}_1^{-1} D_2 + m)^{-1}_{zw} \Delta L_w \>^{(0)} \, . \nonumber
\end{align}
We now write the contribution of $\< R_x^{(2)} \>^{(1)}_{NT}$ and 
$\< R_x^{(3)} \>^{(0)}_{NT}$,
which contains 19 and 12 connected non-tadpole diagrams, respectively:
\begin{align}
g^2 & \< R_x^{(2)} \>^{(1)}_{NT}  = - g^3/\sqrt{2} \{
\< \chi \bar \chi \>_{xz} \< \chi \bar \chi \>_{zu} [ 
\< A_z F_w \> ( \nonumber \\ 
& 4 \< A_u A_w \> D_{2 uv} \< A_v A_w \> + 4 \< B_u B_w \> D_{2 uv} \< B_v B_w \>
\nonumber \\
& - 2 D_{2 uv} \< A_v A_w \> \< A_w A_v \> - 2 D_{2 uv} \< B_v B_w \> \< B_w B_v \> )
\nonumber \\ 
& + \< A_z A_w \> ( 4 \< A_u F_w \> D_{2 uv} \< A_v A_w \>  \nonumber \\
& + 4 \< A_u A_w \> D_{2 uv} \< A_v F_w \> \nonumber \\ 
& - 4 \< B_u G_w \> D_{2 uv} \< B_v B_w \> - 4 \< B_u B_w \> D_{2 uv} \< B_v G_w \>
\nonumber \\ 
& - 2 D_{2 uv} \< A_v F_w \> \< A_w A_v \> + 2 D_{2 uv} \< B_v G_w \> \< B_w B_v \> ) ]
\nonumber \\ 
& + \< \chi \bar \chi \>_{xz} \gamma_5 \< \chi \bar \chi \>_{zu} \gamma_5 [
\< B_z B_w \> ( 4 \< A_u F_w \> D_{2 uv} \< B_v B_w \> \nonumber \\ 
& - 4 \< A_u A_w \> D_{2 uv} \< B_v G_w \> + 4 \< B_u B_w \> D_{2 uv} \< A_v F_w \> 
\nonumber \\ 
& - 4 \< B_u G_w \> D_{2 uv} \< A_v A_w \> - 4 D_{2 uv} \< A_v F_w \> \< B_v B_w \> 
\nonumber \\ 
& + 4 D_{2 uv} \< A_v A_w \> \< B_v G_w \> ) + \< B_z G_w \> ( \nonumber \\
& - 4 \< A_u A_w \> D_{2 uv} \< B_v B_w \> - 4 \< B_u B_w \> D_{2 uv} \< A_v A_w \>  
\nonumber \\ 
& + 4 D_{2 uv} \< A_u A_w \> \< B_v B_w \>)  \} = 0 \, ,
\end{align}

\noindent and
\vspace{-1mm}
\begin{align}
g^3 & \< R_x^{(3)} \>^{(0)}_{NT} = 2 \sqrt{2} g^3 \< \chi \bar \chi \>_{xy}
\{ \< \chi \bar \chi \>_{yz} \< \chi \bar \chi \>_{zw} [ \nonumber \\
& \< A_y A_w \> D_{2 wl} \< A_z A_l \> 
+ \< A_z A_w \> D_{2 wl} \< A_y A_l \>  \nonumber \\ 
& - D_{2 wl} \< A_y A_l \> \< A_z A_l \> ]
- \< \chi \bar \chi \>_{yz} \gamma_5 \< \chi \bar \chi \>_{zw} \gamma_5 [
\nonumber \\ 
& \< A_y A_w \> D_{2 wl} \< B_z B_l \> + \< B_z B_w \> D_{2 wl} \< A_y A_l \>
\nonumber \\ 
& - D_{2 wl} \< A_y A_l \> \< B_z B_l \> ] 
- \gamma_5 \< \chi \bar \chi \>_{yz} \< \chi \bar \chi \>_{zw} \gamma_5 [ 
\nonumber \\ 
& \< A_z A_w \> D_{2 wl} \< B_y B_l \> 
+ \< B_y B_w \> D_{2 wl} \< A_z A_l \>  \nonumber \\
& - D_{2 wl} \< B_y B_l \> \< A_z A_l \> ] 
+ \gamma_5 \< \chi \bar \chi \>_{yz} \gamma_5 \< \chi \bar \chi \>_{zw} [
\nonumber \\ 
& \< B_y B_w \> D_{2 wl} \< B_z B_l \> 
+ \< B_z B_w \> D_{2 wl} \< B_y B_l \> \nonumber \\ 
& - D_{2 wl} \< B_y B_l \> \< B_z B_l \> ]  \} = 0 \, .
\end{align}
\vspace{-1mm}
A similar procedure is used to determined, $\< R_x^{(1)} \>^{(2)}_{NT} = 0 $ 
and $\< F_x \>^{(3)}_{NT} = 0 $. They contain 32 and 47 diagrams, respectively.

\SEC{Concluding Remarks.} 
We showed that the lattice Wess-Zumino model in four dimensions is 
invariant under a deformed supersymmetric transformation to all orders in 
perturbation theory. As a non-trivial check, we performed a two loop
calculation of a one-point WTi associated with this lattice supersymmetry 
and we showed that it is exactly satisfied at fixed lattice spacing. 
This guarantees the restoration of supersymmetry in the continuum limit
without fine tuning. 
Although each term in Eq.~(\ref{wtg3}) vanishes separately (due to the fact that
we are investigating a one-point WTi), the cancellation happens between 
bosons and fermion fields in each term of Eq.~(\ref{wtg3}), as it is required 
in supersymmetry.
Moreover, the expectation value of $R$ is zero.
This result is not in contradiction with the one in Ref.~\cite{Bonini:2005qx}
in which a one-loop (two-point) WTi was investigated and a finite value of 
$\< R \>$ was founded. 
The reason is that one-point WT-identities do not contribute to the renormalization
of the wave function of scalar and fermion fields.

\end{document}